\documentclass[aps,prl,reprint,amsmath,amssymb,longbibliography,nofootinbib,floatfix,superscriptaddress]{revtex4-1}
\usepackage{graphicx}
\usepackage{physics}
\usepackage{xcolor}
\usepackage{comment}
\usepackage{makecell}
\usepackage{hyperref}
\usepackage{bm}
\usepackage{letltxmacro}
\usepackage{mathtools}

\DeclareMathAlphabet\mathbfcal{OMS}{cmsy}{b}{n}

\LetLtxMacro{\originaleqref}{\eqref}

\newcommand{\unit}[1]{\mathbf{\hat{#1}}}

\begin{document}

\title{Acoustic Chirality}
\author{Alex J. Vernon}
\email{alex.vernon@dipc.org}
\affiliation{Donostia International Physics Center (DIPC), Donostia-San Sebasti\'an 20018, Spain}

\author{Konstantin Y. Bliokh}
\email{konstantin.bliokh@dipc.org}
\affiliation{Donostia International Physics Center (DIPC), Donostia-San Sebasti\'an 20018, Spain}
\affiliation{IKERBASQUE, Basque Foundation for Science, Bilbao 48009, Spain}

\begin{abstract}
We reveal a previously unknown continuous symmetry and conservation law in the equations of linear isotropic elasticity, which describe the chirality of elastic waves. We show that the integral chirality is determined by the population imbalance between right- and left-handed transverse phonons, whereas the local chirality density generally involves both transverse and longitudinal wave components. We also introduce the related concepts of acoustic helicity and ``false chirality''. The theory is illustrated with simple interference fields exhibiting distinct distributions of chirality, spin angular momentum, and false chirality. Our results establish chirality as a fundamental property of elastic waves and provide a general theoretical framework for chiral acoustic phenomena.
\end{abstract}

\maketitle

{\it Introduction.---}
Chirality and helicity are closely-related fundamental properties of classical and quantum waves and matter. They play important roles in biology \cite{Palyi_book}, fluid mechanics \cite{Moffatt2014}, relativistic field theory \cite{Peskin_book}, electromagnetism \cite{Lipkin1964, Calkin1965, Afanasiev1996, Trueba1996, Bliokh2011PRA,  Cameron2012, Bliokh2013NJP, Cameron2012_II, Fernandez-Corbaton2013PRL, Alpeggiani2018}, and light-matter interactions \cite{Barron_book, Barron1986CSR, Tang2010, Hendry2010NN, Tang2011, Schaferling2012PRX, Fernandez-Corbaton2012PRA, Bliokh2014PRL, Forbes2021, Ayuso2022PCCP}. According to modern understanding \cite{Barron_book, Barron1986CSR}, the chirality of any object is quantified by a {\it $P$-odd and $T$-even scalar}, where $P$ and $T$ denote the space-inversion (parity) and time-reversal transformations. 
This quantity can either reflect the geometric chirality of classical matter or represent a $P$-odd $T$-even function of a wave field.
For example, in light-matter interactions, the geometric chirality of particles (e.g., molecules) naturally couples to the electromagnetic chirality through chiral dichroism (i.e., differential energy-transfer rate for opposite chiral states) \cite{Tang2010, Hendry2010NN, Tang2011, Schaferling2012PRX, Bliokh2014PRL}. 

In addition, one can employ {\it chiral momentum} transfer, i.e., an optical {\it chiral force}, to separate particles of opposite chirality \cite{Genet2022ACSPhot, Cameron2014NJP, Bliokh2014PRL, Tkachenko2014NC, Kravets2019PRL, Toftul2026RMP}. Since momentum is a $P$-odd and $T$-odd quantity, a chiral momentum or force is characterized by a {\it $P$-even and $T$-odd vector}.

Importantly, electromagnetic chirality and helicity obey fundamental {\it conservation laws} involving the chirality/helicity density and flow \cite{Lipkin1964, Calkin1965, Afanasiev1996, Trueba1996, Bliokh2011PRA,  Cameron2012, Bliokh2013NJP, Cameron2012_II}. These conservation laws are intimately connected, via Noether's theorem, to the {\it dual symmetry} between electric and magnetic fields \cite{Calkin1965, Afanasiev1996, Trueba1996, Bliokh2011PRA,  Cameron2012, Bliokh2013NJP, Cameron2012_II, Fernandez-Corbaton2013PRL, Alpeggiani2018}. In this framework, the chirality/helicity density is a $P$-odd and $T$-even scalar, whereas the chirality/helicity flow density is a $P$-even and $T$-odd vector, closely related to the chiral momentum and {\it spin angular momentum} densities.

Recently, there has been a surge of interest in chiral phenomena in acoustics. First, this is driven by the discovery of {\it chirality-induced spin selectivity}, i.e., spin polarization and transport of electrons in chiral matter \cite{Ray1999S, Gohler2011S, Naaman2019NRC, Evers2022AM}. 
Second, there is a rapidly growing interest in {\it chiral phonons} \cite{Zhang2014PRL, Zhu2018S, Kishine2020PRL, Ishito2023NP, Tauchert2022N, Ueda2023N, Luo2023S, Choi2024NN}, albeit accompanied by some terminological ambiguity between true chirality and spin angular momentum \cite{Juraschek2025NP}. Importantly, unlike the well-established concepts of electromagnetic, fluid-dynamic, and relativistic quantum-particle chirality, acoustics lacks a fundamental quantitative characterization of chirality and helicity in generic inhomogeneous wave fields.

In this work, we fill this gap by introducing acoustic chirality/helicity and their conservation laws for linear elastic waves in isotropic homogeneous solids. Although the elasticity equations, incorporating both longitudinal compression and transverse shear modes, are more complicated and less symmetric than Maxwell's equations in free space, we show that they nevertheless possess an acoustic counterpart of electromagnetic dual symmetry. We also introduce an acoustic version of the so-called ``false chirality'' (a $P$-odd and $T$-odd scalar) \cite{Barron_book, Barron1986CSR}, which can characterize nonreciprocal chiral-like phenomena and is known in electromagnetism as ``magnetoelectric energy'' \cite{Bliokh2014PRL} or ``reactive helciity'' \cite{Nieto-Vesperinas2021PRR}. 

Although isotropic homogeneous solids present the simplest possible acoustic medium, our findings demonstrate that chirality is fundamentally embedded in acoustics at the same level as in electromagnetism. 
Therefore, the chiral quantities and conservation laws derived here provide an important theoretical toolbox for investigating a broad range of acoustic chiral phenomena.  

{\it Acoustic duality and chirality.---}
The acoustic wave field in an isotropic homogeneous elastic medium can be characterized by the local displacement of the medium particles, $\mathbfcal{R} ({\bf r}, t)$, and the corresponding velocity field $\mathbfcal{V} ({\bf r}, t) = \partial \mathbfcal{R} / \partial t$.  
Linear elasticity shares similarities with Maxwell's electromagnetism \cite{Auld_book, Bliokh2025CP}, where the $\mathbfcal{R}$ and $\mathbfcal{V}$ fields can be associated with the magnetic vector-potential and electric field, respectively. 
To further develop this analogy, we introduce an additional vector field $\mathbfcal{F} =  c_t \boldsymbol{\nabla} \times \mathbfcal{R}$ (analogous to the magnetic field, with a minus sign) and a scalar field $\mathcal{G} = c_l \boldsymbol{\nabla} \cdot\mathbfcal{R}$, which characterizes longitudinal sound modes. 
Here, $c_t = \sqrt{\mu/\rho}$ and $c_l = \sqrt{(\lambda+2\mu)/\rho}$ are the velocities of the transverse shear and longitudinal compression modes, respectively, where $\lambda$ and $\mu$ are the Lam\'{e} coefficients and $\rho$ is the mass density of the medium. 

Using the three fields $\mathbfcal{V}$, $\mathbfcal{F}$, and $\mathcal{G}$ (which all share the same velocity dimension), the linear-elasticity wave equation $\partial^2\mathbfcal{R}/\partial t^2 = c_l^2 \boldsymbol{\nabla} (\boldsymbol{\nabla} \cdot \mathbfcal{R}) - c_t^2 \boldsymbol{\nabla}\times (\boldsymbol{\nabla} \times \mathbfcal{R})$ \cite{LL_elasticity, Bliokh2025CP} can be recast as the following system of first-order equations:
\begin{align}
\label{Maxwell}
\frac{\partial\mathbfcal{V}}{\partial t} & =
- c_t \boldsymbol{\nabla}\times\mathbfcal{F}+ c_l \boldsymbol{\nabla} \mathcal{G}\,,\nonumber\\
\frac{\partial \mathbfcal{F}}{\partial t} & = c_t \boldsymbol{\nabla}\times \mathbfcal{V}\,,\nonumber \\
\frac{\partial \mathcal{G}}{\partial t} & = c_l \boldsymbol{\nabla}\cdot\mathbfcal{V}\,.
\end{align}
The field $\mathbfcal{F}$ is purely transverse, $\boldsymbol{\nabla}\cdot \mathbfcal{F} =0$, whereas the velocity field can be decomposed using the Helmholtz decomposition into transverse (divergence-free) and longitudinal (curl-free) components: 
\begin{equation}
\label{helmholtz}
\mathbfcal{V} = \mathbfcal{V}_t + \mathbfcal{V}_l = -c_t\boldsymbol{\nabla}\times\mathbfcal{M} + c_l\boldsymbol{\nabla}\mathcal{K}\,,
\end{equation}
where $\mathbfcal{M}$ and $\mathcal{K}$ are the vector and scalar potentials of the transverse and longitudinal velocity fields $\mathbfcal{V}_t$ and $\mathbfcal{V}_l$, respectively. (Note that $\mathbfcal{M}$ is analogous to the electric vector-potential in electromagnetism \cite{Calkin1965, Afanasiev1996, Trueba1996, Cameron2012, Bliokh2013NJP, Cameron2012_II}.) 

Using the decomposition \eqref{helmholtz}, Eqs.~\eqref{Maxwell} split into two independent subsystems for the fields $(\mathbfcal{V}_t, \mathbfcal{F})$ and $(\mathbfcal{V}_l, \mathcal{G})$. The former describes transverse shear waves and is fully analogous to Maxwell's equations in free space. 
The latter describes longitudinal compression waves and is similar to the sound-wave equations in fluids (where $\mathcal{G}$ parallels the pressure field).
From the similarity of the transverse subsystem to Maxwell's equations, it immediately follows that Eqs.~\eqref{Maxwell} possess a continuous {\it acoustic dual symmetry} parametrized by an angle $\theta$ (cf. \cite{Calkin1965, Afanasiev1996, Trueba1996, Bliokh2011PRA,  Cameron2012, Bliokh2013NJP, Cameron2012_II, Fernandez-Corbaton2013PRL, Alpeggiani2018}):
\begin{align}
\label{dual_transform}
\mathbfcal{V}_t &\to \mathbfcal{V}_t\cos\theta + \mathbfcal{F}\sin\theta\,,\nonumber \\
\mathbfcal{F} &\to \mathbfcal{F}\cos\theta - \mathbfcal{V}_t\sin\theta\,.
\end{align}
%
%
{Equations \eqref{dual_transform} can be interpreted, for each transverse plane wave making up the acoustic field, as a rotation of $\mathbfcal{V}_t$ and $\mathbfcal{F}$ by an angle $\theta$ about the wavevector.}

According to Noether's theorem, the symmetry transformation \eqref{dual_transform} implies the existence of a corresponding conservation law. Indeed, we find the conservation law for {\it acoustic chirality} (cf. \cite{Lipkin1964, Calkin1965, Afanasiev1996, Trueba1996, Bliokh2011PRA,  Cameron2012, Bliokh2013NJP, Cameron2012_II}):
\begin{equation}
\label{continuity_chirality}
\frac{\partial C}{\partial t} + \boldsymbol{\nabla} \cdot \mathbf{U}=0\,,
\end{equation}
where
\begin{align}
\label{chirality_density}
C&=\frac{\rho}{2} \left[\mathbfcal{V} \cdot (\boldsymbol{\nabla}\! \times\! \mathbfcal{V}) + \mathbfcal{F} \cdot (\boldsymbol{\nabla} \!\times\! \mathbfcal{F}) \right] \,  \\
&= \underbracket{\frac{\rho}{2} \left[\mathbfcal{V}_t \cdot (\boldsymbol{\nabla} \!\times\! \mathbfcal{V}_t) + \mathbfcal{F} \cdot (\boldsymbol{\nabla} \!\times \! \mathbfcal{F}) \right]}_{C_t} + \underbracket{\frac{\rho}{2} \boldsymbol{\nabla}\cdot (\mathbfcal{V}_t \!\times\!\mathbfcal{V}_l)}_{C_m} \nonumber
\end{align}
is the chirality density, and 
\begin{equation}
\label{chirality_flux}
{\bf U}\!=\!\underbracket{\frac{\rho}{2}\! \left\{ c_t\! \left[\mathbfcal{F} \!\times\! (\boldsymbol{\nabla} \!\times\! \mathbfcal{V}_t) \!-\! \mathbfcal{V}_t\! \times\! (\boldsymbol{\nabla}\! \times \!\mathbfcal{F})\right] \right\}}_{{\bf U}_t}
+ \underbracket{ \frac{\rho}{2}\frac{\partial}{\partial t }\!(\mathbfcal{V}_l \!\times\!\mathbfcal{V}_t)}_{{\bf U}_m}\,
\end{equation}
is the chirality flow density (defined up to the curl of a vector field). 
One can readily verify that $C$ is a $P$-odd and $T$-even scalar, whereas ${\bf U}$ is a $P$-even and $T$-odd vector. 

Importantly, the chirality density \eqref{chirality_density} and flow density \eqref{chirality_flux} naturally decompose into purely transverse contributions, depending on the transverse fields $\mathbfcal{V}_t$ and $\mathbfcal{F}$ and directly related to the dual symmetry \eqref{dual_transform}, and {\it mixed} contributions involving both transverse and longitudinal fields (cf. the analogous hybrid elastic spin discussed in \cite{Long2018PNAS, Yang2023PRL}). 
These contributions are separately conserved, independently satisfying the continuity equation \eqref{continuity_chirality}.
Although it might seem that the mixed contribution is unnecessary, it is intrinsically embedded in the chirality of the total velocity field described by the $\mathbfcal{V}\cdot  (\boldsymbol{\nabla}\times \mathbfcal{V})$ term, and can therefore contribute to observable chiral phenomena \cite{Yang2023PRL}.
In turn, for purely longitudinal fields, the chirality vanishes identically \cite{Toftul2026RMP}.

The chirality density \eqref{chirality_density} is generally independent of the local spin angular momentum density, which is given by \cite{Bliokh2025CP, Jones1973, Garanin2015PRB, Nakane2018PRB, Long2018PNAS, Shi2019, Yang2023PRL} 
\begin{equation}
\label{spin}
{\bf S}= {\rho} \left( \mathbfcal{R} \times \mathbfcal{V} \right).
\end{equation}
(Note that the spin density itself includes transverse, mixed, and longitudinal contributions \cite{Long2018PNAS, Shi2019, Jones1973, Yang2023PRL}.) 
We also note that a recent paper \cite{Tateishi2025} employed a quantity analogous to $C$ but lacking the $\mathbfcal{F}$-field contribution; however, that quantity does not obey a conservation law. 

Equations \eqref{Maxwell}--\eqref{chirality_flux} constitute the central results of this work. 
To the best of our knowledge, they reveal a previously unidentified symmetry and conservation law in the isotropic elasticity equations. 
Notably, it is the absence of the $\mathbfcal{F}$ field in the standard formulation of elasticity theory that previously prevented the identification of these fundamental properties.

For monochromatic acoustic waves of frequency $\omega$, all fields can be represented in terms of their complex amplitudes: $\mathbfcal{V}({\bf r}, t) = {\rm Re}[{\bf V}({\bf r}) e^{-i\omega t}]$, $\mathbfcal{F}({\bf r}, t) = {\rm Re}[{\bf F}({\bf r}) e^{-i\omega t}]$, and $\mathcal{G}({\bf r}, t) = {\rm Re}[{G}({\bf r}) e^{-i\omega t}]$. 
Substituting these expressions into Eqs.~\eqref{chirality_density} and \eqref{chirality_flux}, using Eqs.~\eqref{Maxwell}, and performing time averaging, we obtain the time-averaged chirality density and flow: 
\begin{align}
\label{chirality_density_mono} 
\bar{C} & = \underbracket{\frac{\rho\omega}{2 c_t} {\rm Im}\!\left({\bf V}_t^* \cdot {\bf F}\right)}_{\bar{C}_t} + \underbracket{\frac{\rho \omega}{4 c_t} {\rm Im}\! \left( {\bf V}_l^* \cdot {\bf F} \right)}_{\bar{C}_m} \,, \\
\label{chirality_flux_mono} 
\bar{\bf U} & = \bar{\bf U}_t = \frac{\rho \omega}{4}{\rm Im}\!\left( {\bf V}_t^*\! \times {\bf V}_t + {\bf F}^*\! \times {\bf F} \right)\,. 
\end{align}
%
The transverse contributions in these equations are fully analogous to their electromagnetic counterparts \cite{Bliokh2011PRA, Cameron2012}. 
In particular, the chirality flow density resembles the transverse part of the spin density $\bar{\bf S} = (\rho/2\omega){\rm Im}({\bf V}^* \times {\bf V})$, but symmetrized with respect to the ${\bf V}_t$ and ${\bf F}$ fields. 

For a transverse plane wave with wavevector ${\bf k}$, one has ${\bf F} = - (c_t/\omega) {\bf k} \times {\bf V}_t$, ${\bf V}_t\cdot {\bf k} =0$, which yields $\bar{C} =\bar{C}_t = \omega {\bf k} \cdot \bar{\bf S}$ and $\bar{\bf U} = \omega^2 \bar{\bf S}$.
Thus, the plane-wave chirality is proportional to the scalar product of the spin and wavevector, as expected for chiral phonons \cite{Juraschek2025NP}. 
Importantly, Eqs.~\eqref{chirality_density} and \eqref{chirality_density_mono} 
describe local acoustic chirality in a much more general case of arbitrarily structured waves, where the plane-wave approximation is no longer applicable (see examples below).

{\it Integral chirality and phonons.---}
We now examine the integral chirality of a localized acoustic wave field: $\langle C\rangle = \int C\, d^3{\bf r}$. 
The mixed part of the chirality density \eqref{chirality_density} vanishes upon integration over all space, so that the integral chirality is determined entirely by the transverse modes: 
$\langle C\rangle = \int C_t\, d^3{\bf r}$. 

To express the integral chirality in terms of phonon amplitudes and numbers, we represent the transverse velocity field as a Fourier integral over plane waves with circular polarizations (helicities) $\sigma = \pm 1$:
\begin{equation}
\label{Vpm}
\mathbfcal{V}_t=\sum_{\sigma}\int\! \frac{d^3\mathbf{k}}{(2\pi)^{3/2}} \sqrt{\frac{\hbar \omega}{2 \rho}} \!\left[ a^\sigma ({\bf k})\unit{e}^\sigma ({\bf k}) e^{i{\bf k}\cdot {\bf r} - i\omega t} + {\rm c.c.}\right] .
\end{equation}
Here, $a^\sigma$ are the amplitudes of right- and left-handed phonons, while $\unit{e}^\sigma$ are the corresponding unit polarization vectors. 
One can show that the integral energy of the transverse field is $\langle E\rangle  = \int \hbar \omega (|a^+|^2 + |a^-|^2)\, d^3{\bf k}$, so that $|a^\pm|^2$ can be regarded as the densities of right- and left-handed phonons in ${\bf k}$-space. 

Using the representation \eqref{Vpm}, the integral acoustic chirality can be calculated in a manner entirely analogous to electromagnetism \cite{Calkin1965, Afanasiev1996, Trueba1996, Bliokh2011PRA, Cameron2012}, yielding
\begin{equation}
\label{chirality_integral}
\langle C\rangle =\int \hbar \omega k (|a^+|^2 - |a^-|^2)\, d^3{\bf k} \,.
\end{equation}
For a quasi-monochromatic field with a narrow frequency spectrum centered around $\omega_0$, Eq.~\eqref{chirality_integral} reduces to $\langle C\rangle \simeq\hbar (\omega^2_0/c_t) (N^+ - N^-)$, where $N^\pm$ are the total numbers of right- and left-handed phonons. 

{\it Acoustic helicity.---}
It is known from electromagnetism that chirality and helicity are closely related quantities \cite{Lipkin1964, Calkin1965, Afanasiev1996, Trueba1996, Bliokh2011PRA,  Cameron2012}. 
For both photons and phonons, the helicity of a plane wave is given by the projection of its spin onto the wavevector direction, while the integral helicity is determined by the difference between the numbers of the right- and left-handed particles:
\begin{equation}
\label{helicity}
\bar{C}_{H} \!= \bar{\bf S}\cdot \frac{\bf k}{k}\,, ~
\langle C_{H} \rangle \!= \!\int\! \hbar (|a^+|^2 - |a^-|^2)\, d^3{\bf k} = \hbar(N^+ - N^-).
\end{equation}
Hence, in the monochromatic limit, the acoustic chirality quantities described by Eqs.~\eqref{continuity_chirality}--\eqref{chirality_flux}, \eqref{chirality_density_mono}, \eqref{chirality_flux_mono}, and \eqref{chirality_integral} should be proportional to the corresponding  helicity quantities, with the proportionality factor $\omega^2/c_t$.

To properly introduce the acoustic helicity density and its flow, satisfying a continuity equation, {we return to the Helmholtz decomposition of the velocity field, Eq.~\eqref{helmholtz}.}
Assuming that $\mathbfcal{M}$ is a purely transverse field, $\boldsymbol{\nabla} \cdot \mathbfcal{M} =0$, we find from Eqs.~\eqref{Maxwell} that $\partial \mathbfcal{M}/\partial t = \mathbfcal{F}$ and $\partial \mathcal{K}/\partial t = \mathcal{G}$. 
Since $\mathbfcal{V} = \partial \mathbfcal{R}/\partial t$, it follows that the fields $\mathbfcal{R}$, $\mathbfcal{M}$, $\mathcal{K}$ satisfy the same equations of motion \eqref{Maxwell} as the fields $\mathbfcal{V}$, $\mathbfcal{F}$, $\mathcal{G}$. Accordingly, they obey the same dual symmetry \eqref{dual_transform} and conservation law \eqref{continuity_chirality}--\eqref{chirality_flux} under the substitution
\begin{equation}
\label{substitution}
\left\{ \mathbfcal{V},~\mathbfcal{F},~\mathcal{G}\right\} \to 
\left\{ \mathbfcal{R},~\mathbfcal{M},~\mathcal{K}\right\}\,.
\end{equation}
Applying this substitution and defining the corresponding densities similar to Eqs.~\eqref{chirality_density}, \eqref{chirality_flux}, \eqref{chirality_density_mono}, \eqref{chirality_flux_mono}, with an additional factor of $c_t$, we obtain the desired helicity density $C_H$ and its flow ${\bf U}_H$, satisfying Eqs.~\eqref{helicity}. 

Notably, unlike the chirality density $C$, the helicity density $C_H$ is gauge-dependent because the vector-potential $\mathbfcal{M}$ {in Eq.~\eqref{helmholtz}} is defined up to the gauge transformation $\mathbfcal{M} \to \mathbfcal{M} + \boldsymbol{\nabla}\varphi$. However, the integral helicity $\langle C_H \rangle$ remains invariant under such gauge transformations.

{\it False chirality.---}
Alongside the proper $P$-odd and $T$-even chirality density, one can also introduce a ``false chirality'' (using the terminology of L.~D.~Barron \cite{Barron_book, Barron1986CSR}) characterized by a $P$-odd and $T$-odd scalar. In light-matter interactions, such a quantity describes the nonreciprocal magnetoelectric effect, i.e., the magnetic (electric) response of matter induced by an external electric (magnetic) field \cite{Bliokh2014PRL}. 
Optical media exhibiting this type of response are also known as ``Tellegen media'' \cite{Tellegen, Tretyakov1998, Ghosh2008, Jazi2024NC, Yang2025NC}.

For the elasticity equations \eqref{Maxwell}, we construct a false-chirality density $D$, satisfying the conservation law $\partial D/\partial t + \boldsymbol{\nabla} \cdot {\bf W} =0$ and expressed through acoustic spin density \eqref{spin}:
\begin{eqnarray}
\label{false_chirality}
&D=\dfrac{\rho}{2}\left[ \mathbfcal{V} \cdot (\boldsymbol{\nabla} \times  \mathbfcal{R}) -\mathbfcal{R} \cdot (\boldsymbol{\nabla} \times  \mathbfcal{V}) \right]
= \dfrac{1}{2} \boldsymbol{\nabla} \cdot {\bf S}\,, \\
\label{false_chirality_flux}
&{\bf W}=\dfrac{\rho}{2}\, \mathbfcal{R} \times \left( 
c_t \boldsymbol{\nabla} \times  \mathbfcal{F} - c_l \boldsymbol{\nabla}G \right)
= - \dfrac{1}{2} \dfrac{\partial {\bf S}}{\partial t}\,.
\end{eqnarray}
The corresponding time-averaged monochromatic quantities are [cf. Eqs.~\eqref{chirality_density_mono} and \eqref{chirality_flux_mono}]:
\begin{align}
\label{false_chirality_mono}
\bar{D} = \frac{\rho}{2 c_t} {\rm Re} ({\bf V}^* \cdot {\bf F})\,, \quad 
\bar {\bf W} = {\bf 0}\,.
\end{align}
The false-chirality density $\bar{D}$ is entirely similar to the corresponding electromagnetic quantities: the magnetoelectric energy density \cite{Bliokh2014PRL} or reactive helicity \cite{Nieto-Vesperinas2021PRR}. 
It vanishes for a plane wave but is generally nonzero in spatially inhomogeneous fields.

{\it Examples.---}
We now present simple examples of non-plane-wave acoustic fields possessing local chirality, spin, and false chirality, which clearly illustrate the distinctions between these quantities. 
The simplest field that cannot be described by a single plane wave is a standing wave formed by the interference of two counter-propagating plane waves with the same frequency $\omega$ and amplitude $A$. Such fields can naturally arise as eigenmodes of finite solid bodies. 

Consider the interference of transverse waves propagating along the $\pm z$ directions with circular polarizations. The corresponding complex wave fields ${\bf V}_t$ and ${\bf F}$ are:
\begin{align}
\label{standing_field}
{\bf V}_t & = A \left( \hat{\bf e}^{\sigma_1} e^{ikz} + \hat{\bf e}^{\sigma_2} e^{-ikz} \right), \nonumber \\
{\bf F} & = A \left(i \sigma_1 \hat{\bf e}^{\sigma_1} e^{ikz} + i\sigma_2 \hat{\bf e}^{\sigma_2} e^{-ikz} \right).
\end{align}
Here, $k=\omega/c_t$ and $\hat{\bf e}^{\sigma_{1,2}} = (\hat{\bf x} \pm i \sigma_{1,2}\hat{\bf y})/\sqrt{2}$ are the circular-polarization unit vectors of the two waves with helicities $\sigma_1$ and $\sigma_2$.

\begin{figure}[t]
\centering
\includegraphics[width=\linewidth]{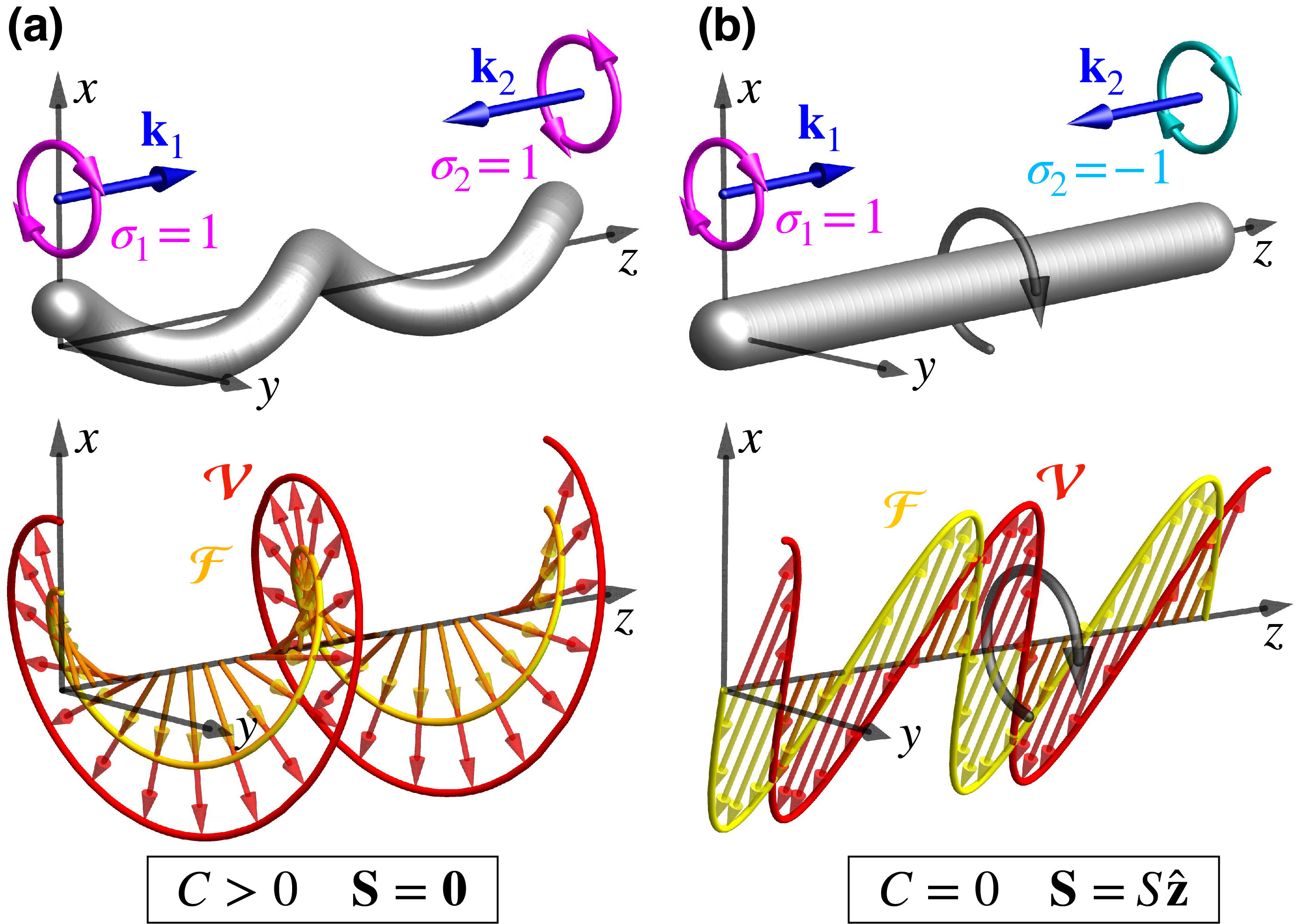}
\caption{(a) Standing transverse wave formed by the superposition of two counter-propagating plane waves with the same helicity. Such a wave can arise as an eigenmode of a helical or twisted rod. The field is linearly polarized at every point, while the polarization directions form a spiral spatial structure (see Supplemental Movie 1). 
This wave possesses nonzero chirality $\bar{C}>0$ and zero spin density $\bar{\bf S}={\bf 0}$. 
(b) Standing transverse wave formed by the superposition of two counter-propagating plane waves with opposite helicities. Such a wave can arise as an eigenmode of a rotating rod. The field is circularly polarized at every point, while the instantaneous field distribution remains planar (see Supplemental Movie 2). This wave has zero chirality $\bar{C}=0$ and nonzero spin $S_z>0$.
The instantaneous distributions of the fields $\mathbfcal{V}(z,t)$ and $\mathbfcal{F}(z,t)$ are shown by the red and yellow arrows, respectively.}
\label{Fig_1}
\end{figure}

The time-averaged energy, momentum, spin, chirality, and false-chirality densities for the field \eqref{standing_field} read:
\begin{align}
\label{standing_energy}
\bar{E} &= \rho A^2\,, ~~ \bar{\bf P} = {\bf 0}\,, ~~
\bar{\bf S} = \frac{\rho A^2}{2\omega}(\sigma_1-\sigma_2)[1+\cos(2kz)]\hat{\bf z} \,, \nonumber \\
\bar{C} &= \frac{\rho A^2k}{2}(\sigma_1+\sigma_2)\,, ~~ \bar{D} = -\frac{\rho A^2}{2c_t}(\sigma_1-\sigma_2)\sin(2kz)\,,
\end{align}
where the energy density can be written as $\bar{E} = \rho \left(|{\bf V}_t|^2 + |{\bf F}|^2 \right)/4$ in this case, and we used the canonical momentum density $\bar{\bf P} = (\rho/2\omega) {\rm Im}[{\bf V}^*\cdot (\boldsymbol{\nabla}){\bf V}]$ \cite{Bliokh2025CP}. Thus, the momentum always vanishes in a standing wave. The chirality is nonzero for waves with the same helicity $\sigma_1=\sigma_2$, while the spin and false chirality are nonzero for waves with opposite helicities $\sigma_1=-\sigma_2$.

Figure~\ref{Fig_1} shows instantaneous distributions of the corresponding real fields $\mathbfcal{V}(z,t)$ and $\mathbfcal{F}(z,t)$ in the standing waves \eqref{standing_field} (see also Supplemental Movies 1 and 2). 
In a `chiral' standing wave ($\sigma_1=\sigma_2$), the fields are {\it linearly}-polarized at every point, while the polarization directions form a {\it spiral} spatial structure. 
By contrast, in a `spinning' standing wave ($\sigma_1= - \sigma_2$), the fields are {\it circularly}-polarized at every point, whereas their instantaneous directions form a {\it planar} structure. (In both cases, $\mathbfcal{V} \parallel \mathbfcal{F}$.)
From symmetry considerations, it is easy to see that such chiral and spinning standing waves can arise as eigenmodes of a helical (or twisted) rod and a rotating rod, respectively. 
These two cases parallel optical chiral (Pasteur) and magneto-active (Faraday) media \cite{Barron_book}.

Let us consider another example, which highlights the {\it mixed} chirality contribution. We examine the interference between an $x$-propagating longitudinal plane wave and a $y$-propagating, $z$-polarized transverse plane wave of the same frequency $\omega$.
The corresponding wave fields in our formalism are:
\begin{align}
\label{TL_field}
{\bf V} & = A \hat{\bf x} e^{ik_l x} + B \hat{\bf z} e^{ik_t y} = {\bf V}_l + {\bf V}_t\,, \nonumber \\
{\bf F}  &= - B \hat{\bf x} e^{ik_t y}\,, \quad
G= -A e^{ik_l x}\,.
\end{align}
Here, $A$ and $B$ are the amplitudes of the two waves, whereas $k_l = \omega/c_l$ and $k_t = \omega/ c_t$ are their wavenumbers.
The time-averaged energy, momentum, spin, chirality, and false-chirality densities of the field \eqref{TL_field} read:
\begin{align}
\bar{E} &= \frac{\rho}{2}\!\left(A^2 + B^2\right) ,~~
\bar{\bf P} = \frac{\rho}{2\omega}\!\left( A^2 k_l\, \hat{\bf x} + B^2 k_t \, \hat{\bf y} \right), \nonumber \\
\bar{\bf S} & = \frac{\rho A B}{\omega}\sin(k_l x - k_t y)\,\hat{\bf y},~~
\bar{C} = \frac{\rho A B k_t}{4} \sin(k_l x - k_t y),
\nonumber \\
\bar{D} & = -\frac{\rho A B}{2 c_t} \cos(k_l x - k_t y)\,.
\end{align}
Here, the chirality density has a purely mixed origin, $\bar{C}=\bar{C}_m$, arising from the interplay between the transverse and longitudinal wave components. The regions of positive and negative chirality alternate throughout the $(x,y)$ plane, so that the spatially-averaged chirality vanishes, $\langle \bar{C} \rangle = 0$, in agreement with Eq.~\eqref{chirality_integral}. 
Nonetheless, the mixed chirality density can manifest itself in local interactions, e.g., with electron or optical fields. 
Note also that the regions of maximal chirality, $k_l x - k_t y = \pi/2 + n\pi$, $n\in \mathbb{Z}$, coincide with the regions of maximal spin density, while the false chirality reaches its extrema in the chirality- and spin-free regions, $k_l x - k_t y = n\pi$.

{\it Conclusion.---}
To summarize, we have uncovered a new conservation law and a continuous symmetry of the linear isotropic elasticity equations, which describe the chirality of elastic waves. Several aspects of acoustic chirality closely parallel the well-established concepts of electromagnetic chirality and the dual (electric-magnetic) symmetry of Maxwell's equations. 
In particular, acoustic chirality involves not only the usual displacement (or velocity) field, but also a second field proportional to the curl of the displacement, which serves as an acoustic analogue of the magnetic field.
An important distinction from electromagnetism is the presence of longitudinal wave modes. Although the integral acoustic chirality is determined solely by the population imbalance between right- and left-handed transverse phonons, the local chirality density has a mixed contribution involving longitudinal acoustic fields. In addition, we have introduced and analyzed the related concepts of acoustic helicity and ``false chirality''. 

We have illustrated the general theory with simple examples of acoustic interference fields exhibiting chirality, spin angular momentum, and false chirality. These examples clearly demonstrate the fundamental differences between these quantities and may be relevant to observable acoustic phenomena.  
In particular, chiral standing acoustic modes in helical solids may underpin the chirality-induced spin selectivity \cite{Ray1999S, Gohler2011S, Naaman2019NRC, Evers2022AM}. 

Our approach can be naturally extended to media with intrinsic chirality, where chiral acoustic responses can be conveniently described through cross-coupling between the $\mathbfcal{V}$ and $\mathbfcal{F}$ fields \cite{Kishine2020PRL}. More broadly, the explicit analogy with Maxwell's electromagnetism, presented in this work, provides a promising framework for future studies of diverse acoustic phenomena.


\begin{acknowledgments}
We acknowledge fruitful discussions with Prof. Jun-ichiro Kishine, Prof. Jorge Puebla, and Prof. Shuichi Murakami.
This work has been co-funded by the European Union through the project HORIZON-MSCA-2022-COFUND-01-SmartBRAIN3-101126600.
\end{acknowledgments}

\bibliography{refs}

\end{document}